\documentclass[%
 reprint,
 amsmath,amssymb,
 aps,
]{revtex4-1}

\usepackage{graphicx}
\usepackage{dcolumn}
\usepackage{bm}
\usepackage{amsmath,amssymb,bbm}
\usepackage[normalem]{ulem}

\usepackage{hyperref,float}
\usepackage[usenames,dvipsnames]{xcolor}
\usepackage{theorem}

\newcommand{\Np}{N_{\rm p}}
\newcommand{\Npt}{N_{{\rm p},t}}
\newcommand{\Nph}{\widehat{N}_{\rm p}}

\DeclareMathOperator*{\argmin}{arg\,min}

\newcommand{\add}[1]{{\color{black}#1}}

{\theorembodyfont{\upshape}
\newtheorem{theo}{Model}}

\begin{document}

\title{Two types of densification scaling in the evolution of temporal networks}

\author{Teruyoshi Kobayashi}
\homepage{Corresponding author: kobayashi@econ.kobe-u.ac.jp}
\affiliation{Department of Economics, Kobe University, Kobe, Japan}
\affiliation{Center for Computational Social Science, Kobe University, Kobe, Japan} 
\author{Mathieu G\'enois}
\affiliation{CNRS, CPT, Aix Marseille Univ, Universit\'e de Toulon, Marseille, France}
\affiliation{GESIS, Leibniz Institute for the Social Sciences, K\"oln/Mannheim, Germany}

\date{\today}%
\vspace*{.5cm}

\begin{abstract}
Many real-world social networks constantly change their global properties over time, such as the number of edges, size and density. While temporal and local properties of social networks have been extensively studied, the origin of their dynamical nature is not yet well understood. Networks may grow or shrink if a) the total population of nodes changes and/or b) the chance of two nodes being connected varies over time. 
Here, we develop a method that allows us to classify the source of time-varying nature of temporal networks. In doing so, we first show empirical evidence that real-world dynamical systems could be categorized into two classes, the difference of which is characterized by the way the number of edges grows with the number of active nodes, i.e., densification scaling. We develop a dynamic hidden-variable model to formally characterize the two dynamical classes. The model is fitted to the empirical data to identify whether the origin of scaling comes from a changing population in the system or shifts in the connecting probabilities.

\end{abstract}

\maketitle
\add{\section{Introduction}}

Along with the increasing availability of high-resolution data sets, dynamics of human social communication have been extensively studied over the past decades~\cite{SocioPatterns,Eagle2006PersUbiquitComput,Onnela2007PNAS,Cattuto2010PlosOne,Stehle2011PLOS,HolmeSaramaki2013book_Springer}. Many of these studies are based on data sets of online interactions, such as emails~\cite{Klimt2004enron}, text messages~\cite{Opsahl2008PhysRevLett,Panzarasa09JASIST} and mobile phones~\cite{Eagle2006PersUbiquitComput,Onnela2007PNAS,Schlapfer2014,ghosh2019quantifying}, but the recent development of sensor devices has also enabled us to collect time-stamped data from face-to-face interactions in physical space~\cite{SocioPatterns,barrat2013empirical,Isella2011JTB,Starnini2013PRL,genois2015data}. Those data therefore cover a wide range of social contexts in which dynamic interactions among individuals form temporal social networks~\cite{HolmeSaramaki2013book_Springer,Masuda2016book}.  

These real-world social networks exhibit very often non stationarity: their structure constantly changes over time not only in shape but also in size. 
Generally, these dynamics are present because the studied system is not closed: it is in fact a common property of real-world social and economic networks that agents are free to enter and exit. 
In online social networks like Twitter and Facebook, anyone can basically join or quit the existing communication at any time. In financial markets, a bank becomes a part of an interbank network if it borrows from or lends to other banks and exits the network when the loan is repaid~\cite{Iori2015,kobayashi2018social}. Another non-conservative aspect of real-world networks is the fact that even if the population is constant, the networking activity might vary due to external factors such as schedule and diurnal rhythm; coffee breaks in a conference~\cite{Isella2011JTB}, pauses between classes in a school~\cite{Stehle2011PLOS,Mastrandrea2015PLOS}, lunch breaks in a company~\cite{genois2015data,genois2018can}, etc.

In the present work, we focus on the evolution of two fundamental quantities that condition the global property of networks: the number of active nodes $N$ and the number of edges $M$. \add{The scaling relationship $M\propto N^\gamma$, known as the ``densification power law," has been found in many real-world systems~\cite{Leskovec2005HepAS,Leskovec2007CA_Full}, where the scaling exponent $\gamma$ is constant and $1<\gamma < 2$. In this work, we present further empirical evidence that there exist two types of scaling in the evolution of networks. In addition to the well-known densification power law, we also show that some systems exhibit an accelerating growth of $M$ in which the scaling exponent itself is increasing.}  

We consider two key factors that would lead networks to be time-varying: $N$ and $M$ will vary over time if a) the size of population (i.e., potential number of active nodes) changes and/or b) the chance of two nodes being connected changes.
Clearly, the size of the population in a system constrains the number of active nodes that form a network: at constant probability for links to appear, more nodes implies more links. Similarly, bilateral matching probability determines the number of edges in the network and thereby its density: at constant population size, the higher the probability for a link to exist, the higher the number of links. The question is then: given an empirical temporal network exhibiting time-varying global quantities, is it possible to identify the source of the dynamics?

Here, we develop a method to perform such a task by exploiting a scaling relationship between the numbers of active nodes and edges. To model the behavior of nodes, we use a dynamical version of a hidden-variable model in which the temporal probability of two nodes being connected is given by a product of ``fitness" parameters~\cite{Caldarelli2002PRL,Boguna2003PRE}. The fitness parameters are considered to be intrinsic and constant features of the nodes. In the present model, the time-evolving aspect arises from two distinct channels. First, we introduce a parameter that modulates the average activity level of nodes. This modulation parameter allows the size of generated networks to vary through a change in the connecting probabilities while keeping the population size, including resting nodes, constant. Second, we allow the population size to vary with time. 
In the original fitness model~\cite{Caldarelli2002PRL,Boguna2003PRE}, there is no distinction between population and the number of active nodes, because the population size is assumed to be large enough so that virtually all nodes in the system are active~\cite{kobayashi2018social}. However, if the population in a system is not sufficiently large, a certain fraction of existing nodes may not be active~\cite{kobayashi2018social}, and thereby a change in the population size affects the rate at which the number of edges grows with the number of active nodes.

In the following, we first expose empirical evidence for the existence of two different types of scaling relationships between $N$ and $M$.
We then present a dynamical hidden-variable model with which we investigate the emergence of the two types of scaling patterns. Specifically, we define two classes of theoretical equations that connect $N$ and $M$ under different specifications on the average activity of nodes and the population size. By identifying a class of equations that better fits the observed data, the proposed method allows us to estimate the actual average activity and the (unobservable) population size. From this we are then able to identify for each empirical data which key factor drives the dynamics of the temporal network. We also briefly mention a variation of the model for cases where the population is fixed and known, allowing us to fit the empirical distribution of node fitnesses to a beta distribution. We conclude by a discussion of our results and the limitations of the model.

\section{Scaling relationship between $N$ and $M$}

\subsection{Data}
\begin{figure*}[t]
    \centering
    \includegraphics[width=16.5cm]{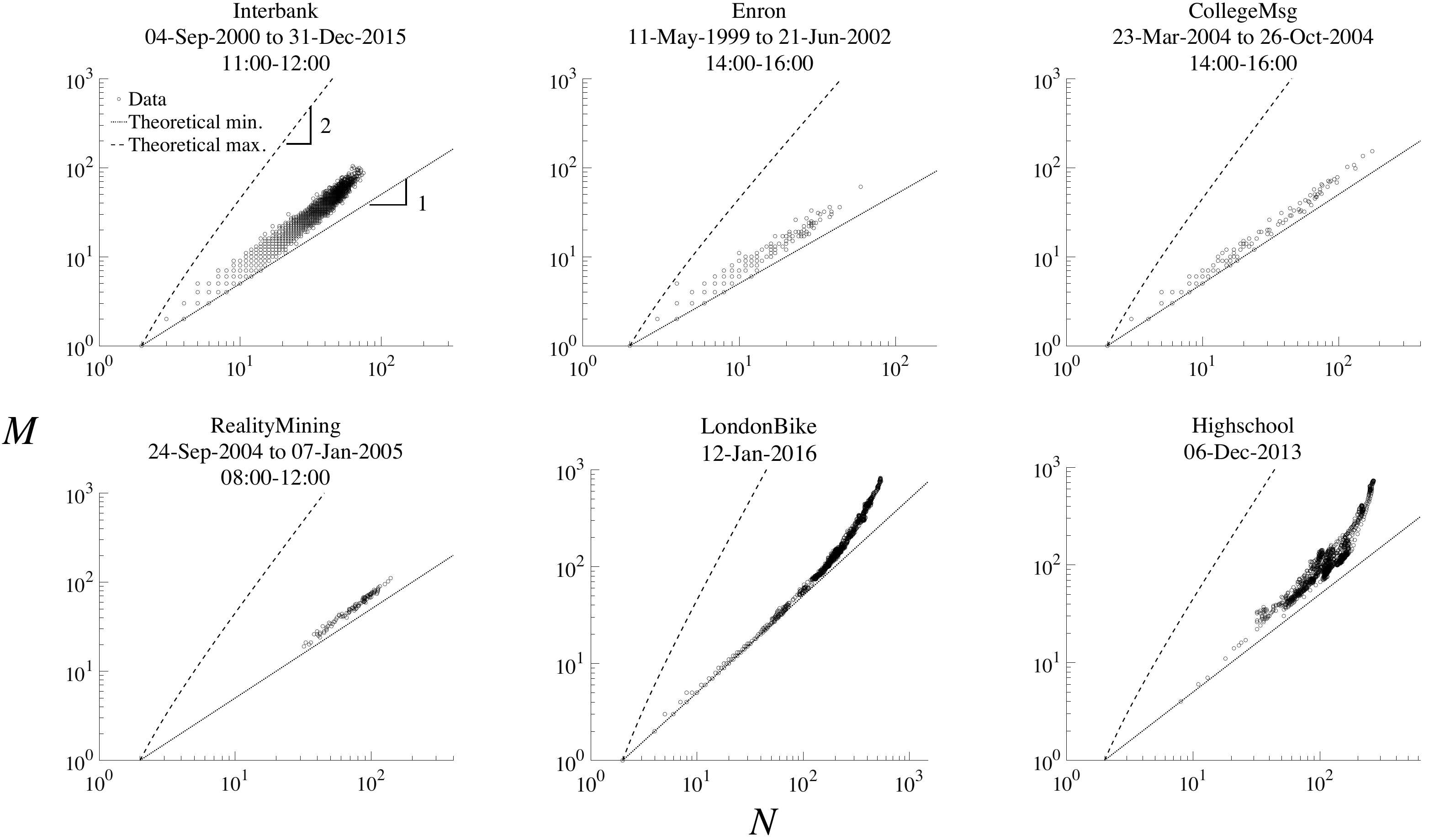}
     \caption{Relationship between the number of active nodes $N$ and the number of edges $M$ in each snapshot for different social contexts. For Interbank, Enron, CollegeMsg and RealityMining data, each dot represents the realization of $(N,M)$ in a particular time window (annotated in the top) of a day. For LondonBike and Highschool data, each dot represents the realization of $(N,M)$ in a 10-minutes time window of a day (0:00--24:00). Black dotted and dashed lines denote the theoretical upper ($M = N(N-1)/2$) and lower ($M = N/2$) bounds, respectively.}
    \label{fig:scaling_data}
\end{figure*}

We consider six data sets of social and economic interest, taken from contexts of very different nature (see Appendix~\ref{sec:method_data} for a full description of the data sets):
\begin{itemize}
    \item Interbank (bilateral transactions in the online interbank market in Italy);
    \item Enron (email communication network from the Enron Corporation~\cite{Klimt2004enron,Leskovec2009enron});
    \item CollegeMsg (online social network at the University of California, Irvine~\cite{Opsahl2008PhysRevLett,Panzarasa09JASIST});
    \item RealityMining (phone call data from the Reality Commons project ~\cite{Eagle2006PersUbiquitComput});
    \item LondonBike (bike trips from the London Bicycle Sharing Scheme~\cite{LondonBike});
    \item Highschool (face-to-face contacts network in a French high school~\cite{Mastrandrea2015PLOS}).
\end{itemize}
All data sets are converted to temporal networks with undirected and unweighted edges. Bidirectional edges (i.e., edges in both directions) are regarded as undirected edges with weight 1. From each dataset, we construct a sequence of network snapshots by defining particular time intervals in each of which all the interactions between nodes are regarded as the edges of the corresponding network. \add{We define $N$ to be the number of nodes that have at least one edge in a given snapshot, and $M$ denotes the corresponding number of edges. This suggests that $N/2 \leq M \leq N(N-1)/2$, where $N/2$ corresponds to the minimum number of edges that \add{can exist between $N$ active nodes} (when all the nodes \add{are connected to}exactly one edge), while $N(N-1)/2$ corresponds to \add{the maximum number of edges that can exist between $N$ active nodes} (i.e., complete graph).}

\subsection{Evidence from empirical data}

We investigate the dynamical relationship between $N$ and $M$. Fig.~\ref{fig:scaling_data} shows scatter plots of $M$ against $N$ for each social context. Two important features appear. First and foremost, there is a strong positive correlation between $N$ and $M$ in all the data sets we examine. \add{In particular, we observe superlinear scaling, \emph{i.e.}, the rate at which $M$ rises with $N$ is larger than that expected by a linear growth, as is occasionally reported for many real-world systems~\cite{Leskovec2005HepAS,Leskovec2007CA_Full}. This phenomenon is also known as the ``densification power law" or  ``densification scaling"~\cite{Leskovec2005HepAS,Leskovec2007CA_Full,bettencourt2009scientific}.} 

Second, there are two different patterns as to how $M$ grows with $N$. One is the \add{densification scaling} we mentioned, in which the scaling exponent is constant ($>1$), showing as a straight line on a log-log scale plot. In Fig.~\ref{fig:scaling_data}, Interbank, Enron, CollegeMsg and RealityMining appear to belong to this category.
Contrarily, for LondonBike and Highschool \add{the growth of $M$ for large values of $N$ is accelerating: the slope itself increases in log-log space as $N$ grows}~\cite{dorogovtsev2001accelerating}.

Both behaviors are striking, as they suggest the existence of simple mechanisms for the dynamics of global activity in temporal networks.
However, the empirical dynamical relationship we observe in Fig.~\ref{fig:scaling_data} cannot be reproduced by a class of common growing network models in which a new node joins the network with a given number of edges~\cite{Barabasi1999Science,dorogovtsev2000structure,krapivsky2001organization}. While these models are intended to explain the emergence of scaling in empirical degree distributions~\cite{Barabasi1999Science},  the number of edges has a linear correlation with the number of active nodes asymptotically, i.e., $M \propto N$, which is not consistent with our finding.
In the following, we present a model which explains how these different types of global behaviors can emerge from temporal social interactions.

\section{A dynamic hidden-variable model}
 
 \subsection{Model}
To explain the two types of scaling in a unified model, we consider a dynamical version of the hidden variable model in which the probability of two nodes being connected at time interval $t$ is given by:
\begin{align}
    p_{ij,t} = \kappa_t  a_{i} a_{j}, \;\;\; i,j=1,\ldots, \Npt, \;\; t = 1,\ldots, T.
    \label{eq:prob_ij}
\end{align}
where $a_{i}$ is the ``fitness'' of node $i$ that represents the activity level of the node~\cite{Caldarelli2002PRL,Boguna2003PRE,DeMasi2006PRE,Starnini2013PhysRevE}. In the baseline model we assume that $a_i$ is uniformly distributed on $[0,1]$ because in general we do not have any prior information about the distribution of activity levels. We will also consider a beta distribution as an alternative case in section~\ref{sec:betadist}.

There are two time-varying parameters in the model. One is $\Npt$ which represents the potential number of active nodes in the system at time $t$, \emph{i.e.}, the total of active and inactive nodes that are in the system at time $t$. The number of active nodes having at least one edge at time $t$ is denoted by $N_t$. We note that the number of active nodes $N_t$ is always observable, but the potential number of nodes $\Npt$ is not. In social networks, for instance, we do not usually know how many people are ready to interact with other people and what fraction of them actually created at least one edge. In many cases, what we can observe from data is the number of active nodes that appear in the record of interaction history, while there is no record of nodes without interactions. Since the observed active nodes may account for only a fraction of the potential nodes, it is generally written as $N_t = (1-q_{0,t})\Npt$, where $q_{0,t}$ denotes the fraction of inactive nodes that have no edge at time $t$, or equivalently, the probability of a randomly chosen node being isolated.
To take an example of social networks, changes in $\Np$ may represent a situation in which the number of students in the classroom changes over time according to the class schedule, leading to a variation in the maximum possible size of face-to-face contact networks.
The potential number of nodes that are ready to interact with others is the first key parameter of the model, as it physically constrains the size of networks to be observed.

The second time-varying parameter of the model is $\kappa_{t}>0$, which modulates the global activity level of nodes. In the financial system, for instance, the chance that two banks trade during the lunch time would be intrinsically lower than that in the morning~\cite{kobayashi2018social}, in which case the banks' activity levels may have a certain diurnal pattern. In social networks where individuals communicate with each other, $\kappa$ would vary according to the time-schedule of the school, workplace, academic conferences, or the circadian rhythm of humans~\cite{Cattuto2010PlosOne,Jo2012NewJPhys,aledavood2015digital,kobayashi2019structured}.

With this specification, the observed network size $N$ and the number of edges $M$ co-evolve as either $\Np$ or $\kappa$ or both change over time.
One can see a change in $N$ due to a shift in $\Np$ represents an \emph{extensive margin effect}, while a shift in $\kappa$ leads to an \emph{intensive margin effect}. Parameters $\kappa$ and $\Np$ can thus explain two different origins of the time-varying nature of networks.

\add{\subsection{Analytical expression for $N$ and $M$}}

In this section we show an analytical solution for the dynamic hidden-variable model when the network size is not necessarily large enough~\cite{kobayashi2018social}.
Suppose that node~$i$ $(1 \leq i \leq \Np)$ is assigned activity $a_i \in [0,1]$ which is drawn from density $\rho(a)$~\cite{Caldarelli2002PRL}.
The numbers of active nodes $N$ and edges $M$ can be expressed as functions of parameters $\kappa$ and $\Np$ (We drop time subscript $t$ for notational convenience):
\begin{align}
\begin{cases}
    N &= (1- q_0(\kappa,\Np)) \Np,\\
    M &= \frac{\overline{k}(\kappa, \Np) \Np}{2},
\end{cases}
\label{eq:NM_main}
\end{align}
where $\overline{k}(\kappa,\Np)$ denotes the average degree over all the existing nodes including isolated ones. 
To obtain the functional forms of $N$ and $M$, we need to find the functional forms of $q_0(\kappa,\Np)$ and $\overline{k}(\kappa,\Np)$.

\begin{figure}[thb]
    \centering
    \includegraphics[width=7.5cm]{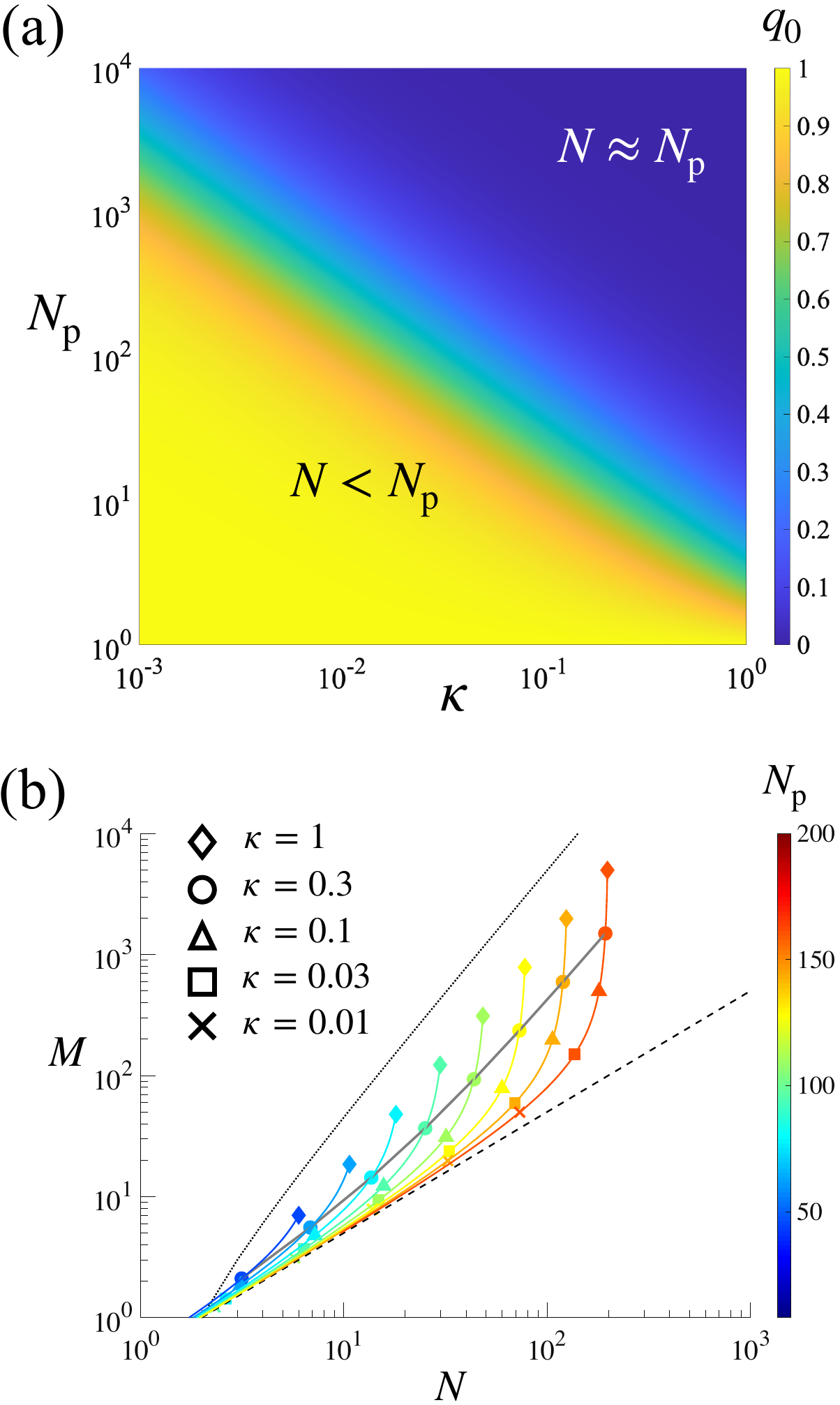}
    \caption{\add{Scaling in the dynamic hidden variable model. (a) Fraction of inactive nodes in the population, $q_0$. (b) Two types of scaling relationships between $N$ and $M$. Each color represents a particular value of $\Np$, while different symbols denote different values of $\kappa$. Each colored line represents a scaling relation when $\Np$ is fixed at a given value, i.e., scaling driven by time-varying $\kappa$. Gray solid line represents a scaling relation when $\kappa$ is fixed ($ = 0.3$), i.e., scaling driven by time-varying $\Np$.}}
    \label{fig:scaling_theory}
\end{figure}

Let $u(a,a^\prime)$ be the probability that there is an edge between two nodes having activity levels $a$ and $a^\prime$, respectively.  As is shown in Appendix~\ref{sec:analytical}, the average degree $\overline{k}(\kappa,\Np)$ is given by:
\begin{align}
    \overline{k}(\kappa,\Np) = (\Np-1) \int \int d a d a^\prime \rho(a)  \rho(a^\prime) u(a, a^\prime),
    \label{eq:k_avg}
\end{align}
which simply states that the average degree is equal to the number of nodes (excluding the focal node itself) times the expected connecting probability. 
It should be noted that Eq.~\eqref{eq:k_avg} is equivalent to Eq.~(21) of Ref.~\cite{Boguna2003PRE} if $\Np -1$ is replaced with $N$, which is asymptotically true as will be shown below.
 From \eqref{eq:q_0} in Appendix~\ref{sec:analytical}, the probability of a randomly chosen node being isolated is given by:
\begin{align}
    q_0(\kappa,\Np) 
    = \int d a^\prime \rho(a^\prime) \left[ 1 - \int u(a^\prime, a) \rho(a) d a \right]^{\Np-1}.
    \label{eq:q_0}
\end{align}
Substituting $\rho(a) = 1$ (i.e., uniform distribution on $[0,1]$) and $u(a, a^\prime) = \kappa a a^\prime$ into Eq.~(\ref{eq:k_avg}) gives:
\begin{align}
    \overline{k}(\kappa,\Np) 
    = \frac{\kappa }{4} (\Np-1).
\end{align}
Similarly, $q_0$ is given by
\begin{align}
    q_0(\kappa,\Np) &= \int_0^1   \left( 1 - \frac{\kappa a^\prime}{2}  \right)^{\Np-1}d a^\prime.
\end{align}
By defining a variable $x \equiv 1 - \frac{\kappa a^\prime}{2}$, we have:
\begin{align}
    q_0(\kappa,\Np) &=  \frac{2}{\kappa}\int_{1-\frac{\kappa}{2}}^1  x^{\Np-1} dx  \notag \\
    &= \frac{2}{\kappa\Np}\left[1-\left( 1-\frac{\kappa}{2}\right)^{\Np}\right].
\end{align}
Combining these results with Eq.~(\ref{eq:NM_main}), we have:
\begin{align}
N &= \Np \left[ 1-  \frac{2}{\kappa\Np}\left(1-\left( 1-\frac{\kappa}{2}\right)^{\Np}\right) \right],\label{eq:N}\\
M &= \frac{1}{8} \kappa\Np(\Np-1).\label{eq:M}
\end{align}
This leads to interesting limit behaviors: if $|1-\kappa/2| < 1$ and $\Np$ is sufficiently large, then $q_0(\kappa,\Np) \simeq 0$ and thereby $N \simeq \Np$ and $M \propto N^2$, as is shown in the study of the static fitness model~\cite{Caldarelli2002PRL,Boguna2003PRE,DeMasi2006PRE}. In contrast, if $\Np$ is not large enough, then $q_0(\kappa,\Np) > 0$ and $N < \Np$, in which case $M$ is not of order $N^2$ and the scaling exponent will take a value between 1 and 2 as is observed in empirical data (Fig.~\ref{fig:scaling_data})~\cite{kobayashi2018social}. Note that $\kappa$ is not \emph{per se} a probability, and that its value does not have any \emph{a priori} upper bound (as it depends on the activity distribution). Clearly, the larger the population $\Np$ and the overall activity $\kappa$, the lower the share of resting nodes $q_0$ in the population (Fig.~\ref{fig:scaling_theory}a).

\subsection{Role of $\kappa$ and $\Np$ in the emergence of scaling}

Using Eqs.~\ref{eq:N} and \ref{eq:M}, we are now able to analyze numerically how $M$ scales with $N$. First, we observe that if the value of $\kappa$ is kept constant while $\Np$ varies, the dynamical relationship between $N$ and $M$ is close to a straight line in a log-log plot (gray solid in Fig.~\ref{fig:scaling_theory}b), as seen in some empirical data. \add{If $\kappa$ is small enough, the scaling is close to linear, approaching the lower bound of $M$ indicated by the dashed line in Fig.~\ref{fig:scaling_theory}b}. However, as $\kappa$ increases, the scaling becomes more and more superlinear, which can be seen in Fig.~\ref{fig:scaling_theory}b by following the same symbols in different colors. 

By contrast, if we vary $\kappa$ for a given value of $\Np$, the slope will bend upward. This can be seen in Fig.~\ref{fig:scaling_theory}b by following different symbols in the same color (colored line). This reproduces the accelerating growth behavior observed in the empirical data, namely LondonBike and Highschool.
\add{We note that although the scaling relationships appear to be quite regular, it proves to be very difficult (if not impossible) to extract from Eqs.~\ref{eq:N} and \ref{eq:M} an analytical expression for them, because of the complicated dependencies of $N$ on $\Np$.}

\section{Identifying the source of network dynamics}

\subsection{Estimation of model parameters from empirical data}\label{sec:method_estimation}

We now propose a method to identify the dynamical class of a system and at the same time estimate $\kappa$ and $\Np$ from the empirical data. In fact, the two model parameters may be estimated in two ways. One is to directly solve the two nonlinear equations Eqs.~\eqref{eq:N} and \eqref{eq:M} with respect to ($\kappa,\Np$) for a given observation of $(N,M)$. This direct calculation gives us a one-to-one mapping of $(N,M)$ to $(\kappa^*,\Np^*)$, where asterisk denotes the solution of the system of two equations. However, such a method proves to be unable to accurately estimate the parameters when the network is small, where there is a large degree of overlap of $(N,M)$ generated under multiple combinations of $(\kappa,\Np)$ (Figs.~\ref{fig:scaling_theory}b and \ref{fig:scaling_validation}a).

\subsection{Exploiting the dynamical relationship between $N$ and $M$}

\subsubsection{Two classes of models for the two types of scaling}

The other method is to use the dynamical relationship between $N$ and $M$. 
This method is based on the idea that the estimation bias due to the overlap of $(N,M)$ could be avoided if we exploit the dynamical relationship between $N$ and $M$ rather than a particular \add{observation} of $(N,M)$ in a given snapshot. In this method, we fit the empirical $N$--$M$ relationship to theoretical equations, which will give us nonlinear least squares estimators of $\kappa$ and $\Np$. 

Since the observable variables $N$ and $M$ appear separately in Eqs.~\eqref{eq:N} and \eqref{eq:M}, respectively, we formulate a regression equation by relating $N$ with $M$ through the substitution of $\kappa$ or $\Np$.
By doing so, we essentially categorize the empirical dynamic networks into two classes. In the regression equation for the first class, we express $N$ as a function of $M$ and parameter $\kappa$ to endogenize the time-variation of $\Np$. Hereafter we call this type of formulation ``Model I". This corresponds to a situation in which the connecting probability $p_{ij}, \forall i\neq j$ is constant while the potential size of networks $\Np$ is time-varying.

In ``Model II", on the other hand, we specify $N$ as a function of $M$ and parameter $\Np$ to endogenize the time-variation of $\kappa$. This type of model would be appropriate when the set of nodes is fixed while the connecting probabilities are affected by diurnal or circadian rhythms.

The regression equations in the two models are respectively given as follows:
\begin{theo} {\emph{$\Np$ is time-varying and $\kappa$ is constant.}}
\begin{align}
    N &= G(M;\kappa) \notag \\
      & \equiv \Np(M,\kappa) \left[ 1-  \frac{2}{\kappa\Np(M,\kappa)}\left(1-\left( 1-\frac{\kappa}{2}\right)^{\Np(M,\kappa)}\right) \right],\label{eq:model1}
\end{align}
where $\Np$ is expressed as a function of $M$ and $\kappa$: $\Np(M,\kappa) \equiv \frac{1+ \sqrt{1+{32M/{\kappa}}}}{2}$ (Eq.~\ref{eq:M}).
We obtain the estimator of $\kappa$, denoted by $\widehat\kappa$, by regressing $N$ on $M$ using a method of nonlinear least squares, where $\bm{N}=G(\bm{M};\widehat\kappa) + \bm{\varepsilon}_{\rm I}$, and $\bm{\varepsilon}_{\rm I}$ denotes the $T\times 1$ vector of residuals. 
Estimates of time-varying $\Np$ are then given by:
\begin{equation}
    \widehat N_{{\rm p},t} = \frac{1+ \sqrt{1+{32M_{t}/{\widehat\kappa}}}}{2}.
\end{equation}
\end{theo}

\begin{theo} {\emph{$\Np$ is constant and $\kappa$ is time-varying}}.
\begin{align}
    N &= F(M;\Np) \notag \\
      & \equiv \Np \left[ 1-  \frac{2}{\kappa(M,\Np)\Np}\left(1-\left( 1-\frac{\kappa(M,\Np)}{2}\right)^{\Np}\right) \right],\label{eq:model2}
\end{align}
where $\kappa$ is expressed as a function of $M$ and $\Np$: $\kappa(M,\Np) \equiv \frac{8M}{\Np(\Np-1)}$ (Eq.~\ref{eq:M}).
We estimate $\Nph$ based on a nonlinear regression equation $\bm{N}=F(\bm{M};\Nph) + \bm{\varepsilon}_{\rm II}$. Estimates of time-varying $\kappa$ are given by:
\begin{equation}
    \widehat\kappa_t = \frac{8M_t}{\Nph(\Nph-1)}.
\end{equation}
\end{theo}

\begin{figure}[tb]
    \centering
    \includegraphics[width=8.5cm]{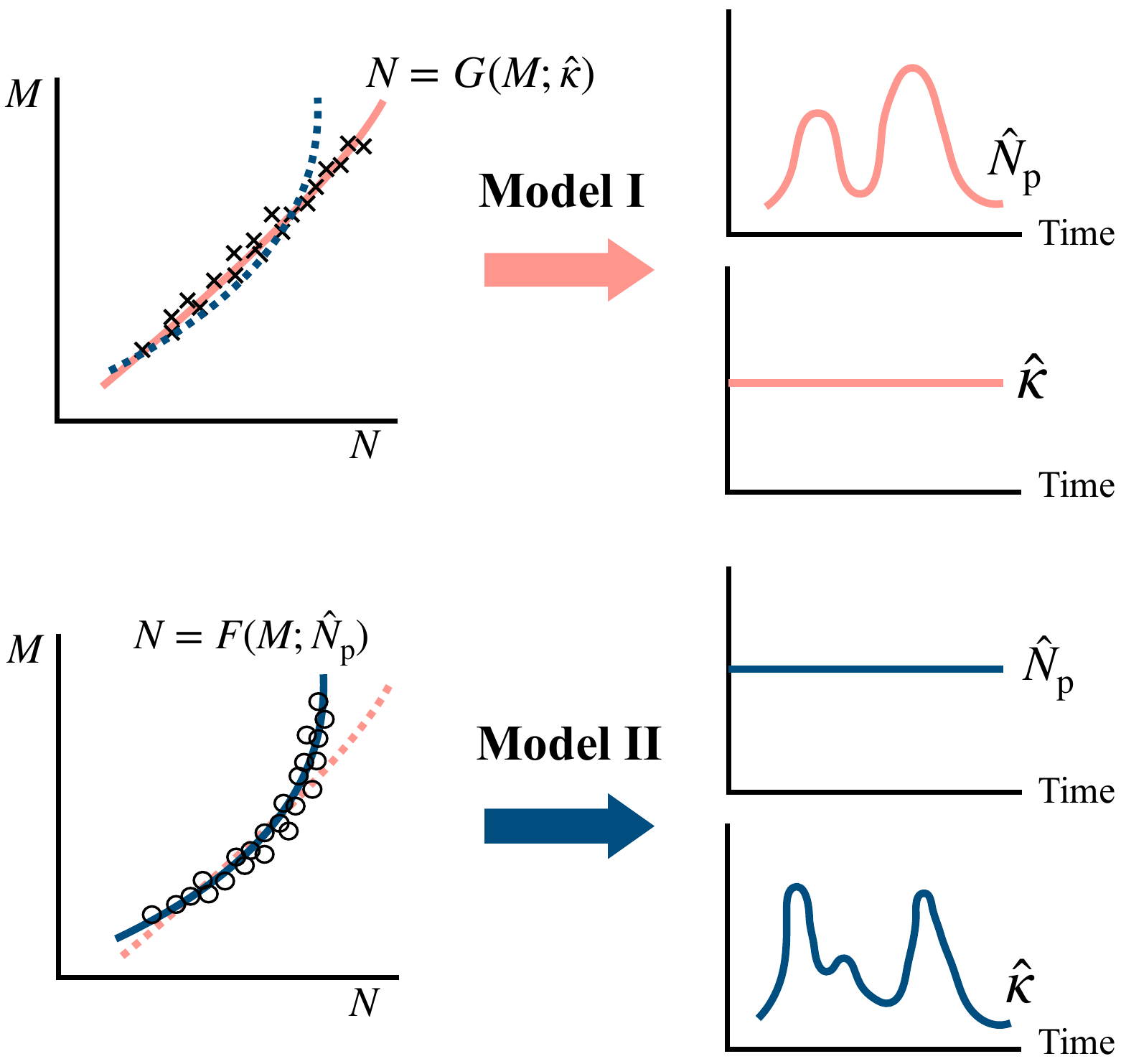}
    \caption{Schematic of model selection. If Model I (resp. Model II) is selected, each bilateral connection probability $\hat{\kappa}$ (resp. population size $\Nph$) is constant while $\Nph$ (resp. $\hat{\kappa}$) is time-varying.}
    \label{fig:schematic}
\end{figure}

After estimating the parameters in both specifications, we select one that attains the lower sum of squared errors: 
\begin{align*}
\begin{cases}
    \text{ Model I is selected if } \bm{\varepsilon}_{\rm I}^\top\bm{\varepsilon}_{\rm I}<\bm{\varepsilon}_{\rm II}^\top\bm{\varepsilon}_{\rm II}, \\
    \text{ Model II is selected otherwise.}
\end{cases}
\end{align*}
A schematic of the model selection is illustrated in Fig.~\ref{fig:schematic}.
Note that the criterion of model selection is effectively the same as that of the Akaike information criterion (AIC) and Bayesian information criterion (BIC) because we have only one parameter in both models.

\subsubsection{Validation}\label{sec:method_validation}

We check the accuracy of the proposed estimation method by using synthetic networks.
For the estimation of Model I (resp. Model II), we generate 500 synthetic networks under various $\Np$ ranging from 20 to 300 (resp. $\kappa$ ranging from 0.001 to 0.99) for a given $\kappa$ (resp. $\Np$).

\begin{figure*}[thb]
    \centering
    \includegraphics[width=14cm]{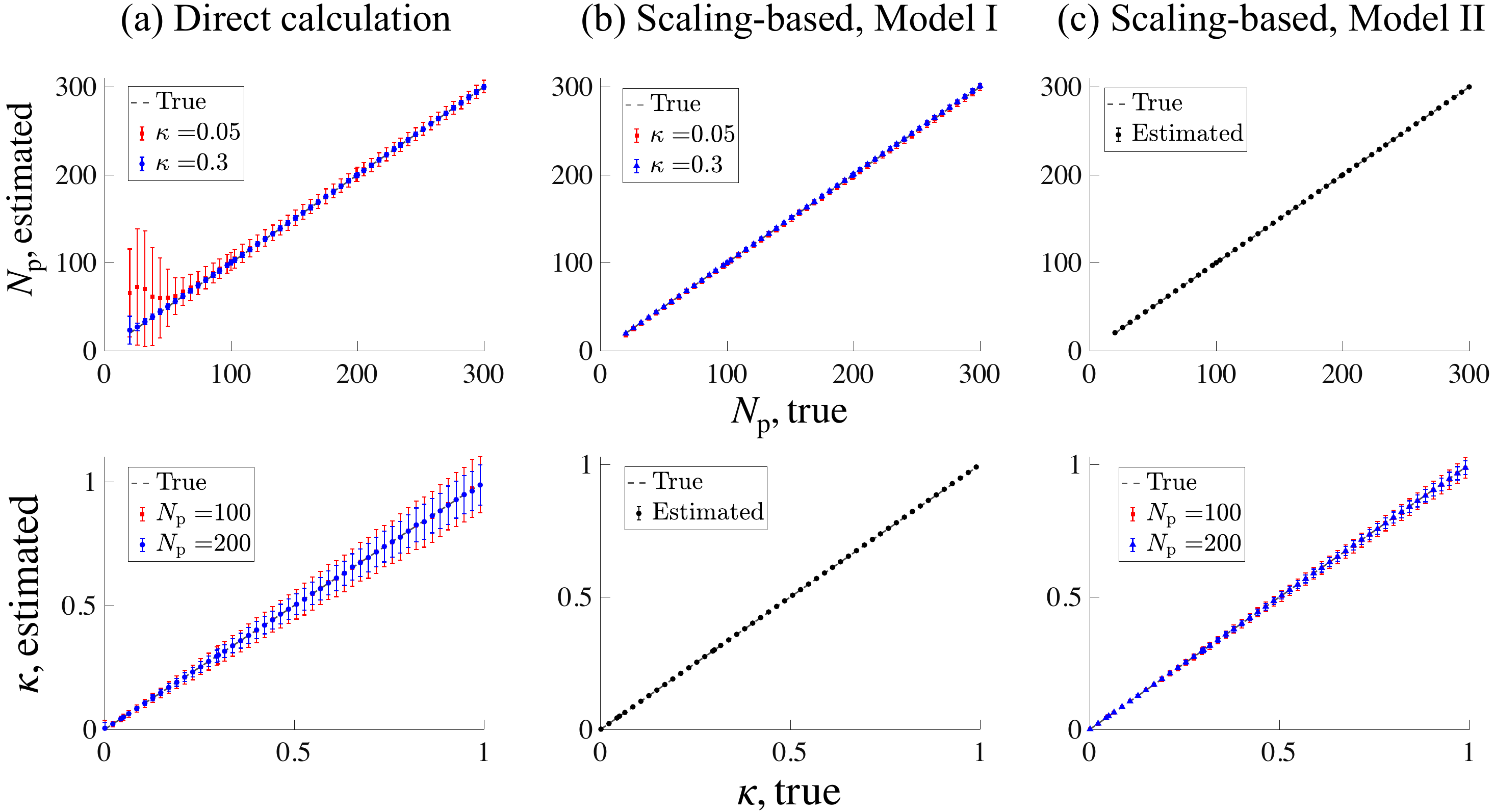}
     \caption{Validation of estimation methods. Error bars are calculated over 1,000 runs. (a) Solutions of Eqs.~\eqref{eq:N} and \eqref{eq:M}. (b) Estimation for Model I, in which $\Np$ is time-varying and $\kappa$ is constant. (c) Estimation for Model II, in which $\Np$ is constant and $\kappa$ is time-varying.}
    \label{fig:scaling_validation}
\end{figure*}

While solving the system of two nonlinear equations is straightforward in principle, the question is whether the obtained solution matches the true values of $\kappa$ and $\Np$. Obviously, the network generating mechanism is in reality not deterministic but stochastic, which means the same parameter combination $(\kappa,\Np)$ may yield different observations of $(N,M)$. Using a particular pair of $(N,M)$ is therefore not sufficient to infer the true model parameters. Indeed, the solution of Eqs.~\eqref{eq:N} and \eqref{eq:M} leads to a biased estimate of $\Np$ especially when the true values of $\kappa$ and $\Np$ are small (Fig.~\ref{fig:scaling_validation}a). This is expected from Fig.~\ref{fig:scaling_theory}b in which there is a large amount of data overlap in the lower left area of the corn. In fact, $\kappa$ tends to take small values (e.g., $< 0.1$) in real-world networks, in which case the biased estimation can become a serious problem.

Fig.~\ref{fig:scaling_validation}b and c shows the error bars of the estimated parameters for the second method over 1,000 runs. The estimated values of $\Np$ and $\kappa$ nicely match the true values even when the network size is fairly small and thereby multiple combinations of $(\kappa,\Np)$ can yield the same $(N,M)$. This is an advantage of this method with which we do not rely on a particular realization of $(N,M)$, but rather we exploit the whole dynamical relationship. Furthermore, in the case where $\Np$ is fixed and $\kappa$ varies in time, Model II also gives a better estimate than the direct calculation.

It should be noted that the observed $N$ can be much lower than its potential value $\Np$, which suggests that the potential number of active nodes cannot necessarily be inferred directly from the observed number of nodes. This is particularly true when $\kappa$ is so small that the network is fairly sparse (Fig.~S1 in SI).

\subsubsection{Empirical results}


\begin{figure*}[thb]
    \centering
    \includegraphics[width=15cm]{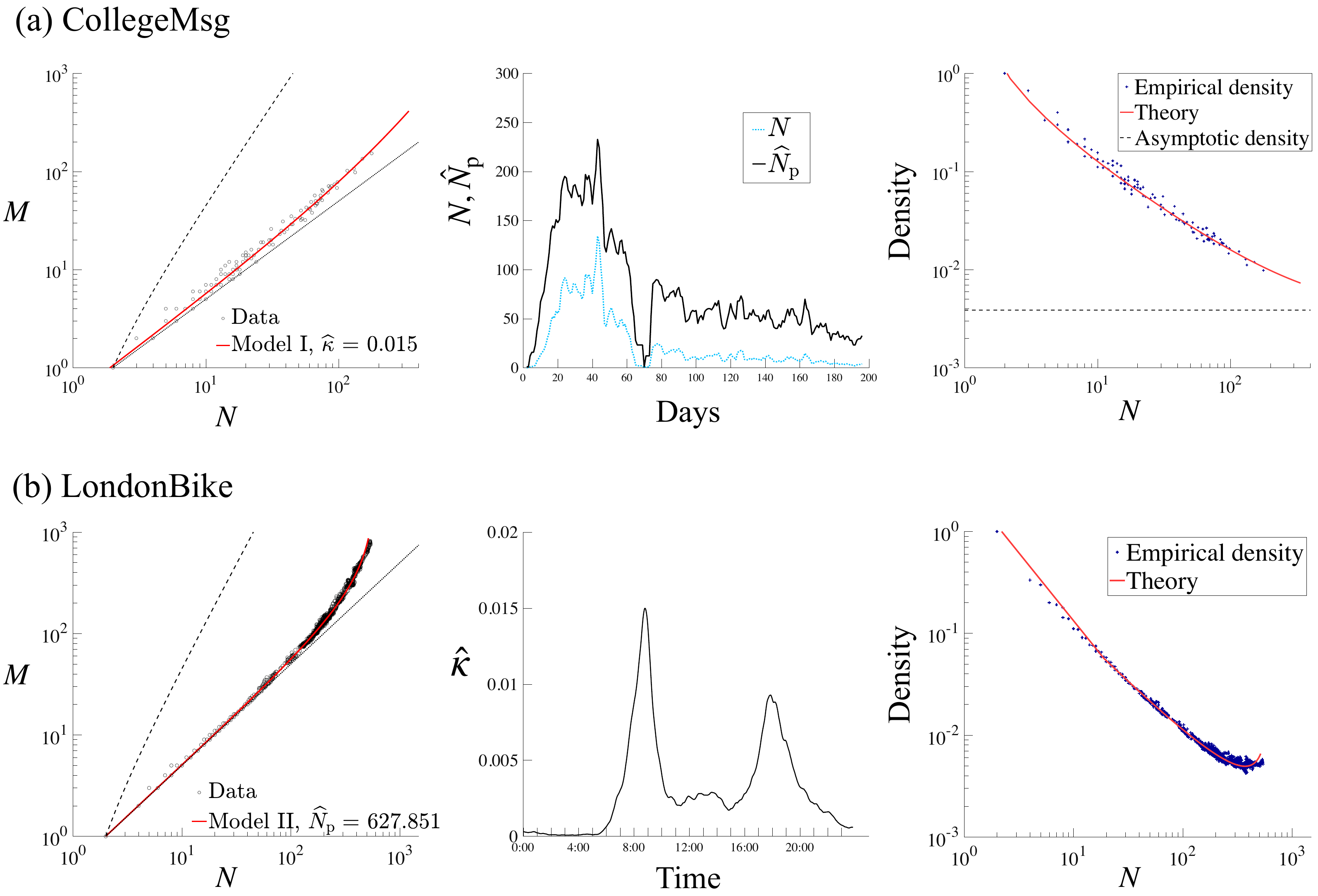}
    \caption{Estimation results for (a) CollegeMsg and (b) LondonBike data sets. Model I is selected for CollegeMsg and Model II is selected for LondonBike. In each panel, left column shows the fitted $N$--$M$ curves. Black dotted and dashed lines denote the theoretical upper ($M = N(N-1)/2$) and lower ($M = N/2$) bounds, respectively. Middle column shows in (a) the estimated $\Nph$, in (b) the estimated $\hat{\kappa}$. Right column shows the fitted theoretical density. In (a), the asymptotic density derived from Eq.~\eqref{eq:density} is marked by the dotted line.}
    \label{fig:scaling_fit}
\end{figure*}

Fig.~\ref{fig:scaling_fit} shows the empirical results for the CollegeMsg and LondonBike data sets (see Figs.S2 and S3 in SI for the other data sets).
\add{Our results illustrate the fact that scaling relations in social and economic temporal networks may be driven by the two previously described factors. For Interbank, Enron, CollegeMsg and RealityMining, Model I is selected, which means the time-varying nature of the global network properties comes from shifts in the potential number of nodes, \emph{i.e.} the population in the system changes over time. On the other hand, for LondonBike and Highschool, Model II is selected, which means the population remains almost unchanged, and the changes in the numbers of edges and active nodes are due to time-varying connecting probabilities. }

Since all we need for the model classification is a variety of combinations of $(N,M)$, one can implement the method for any timescale. For instance, if we see intra-day activity in the Interbank data set, the scaling behaviors on the vast majority of days are still better explained by Model I (Fig.~S4). For LondonBike dataset, the scaling relationship over different days is identified as being driven by a time-varying $\kappa$ for each time interval (Fig.~S5), again indicating that the population (i.e., the number of bike stations) is essentially fixed throughout the data period.
  
For the data sets for which Model I is selected, we note that the estimated values of $\kappa$ are fairly small, ranging from 0.011 (RealityMining) to 0.078 (Interbank). $\widehat\kappa$ is time-varying for LondonBike and Highschool, but their values are still small with the maximum value being no larger than $0.02$. This suggests that the direct calculation discussed above would not work well for empirical networks (Fig.~\ref{fig:scaling_validation}a).

\subsection{Network density}\label{sec:method_density}
  
 Another global quantity that might be of interest is network density.
From the estimates of $\kappa$ and $\Np$ we can write the theoretical network density as
\begin{align}
    \frac{2M}{N(N-1)} = \frac{\widehat\kappa}{4} \left( \frac{1}{1-q_0(\widehat\kappa,\widehat{N}_{\rm p})}\right)^2 \left( 1+\frac{q_0(\widehat\kappa,\widehat{N}_{\rm p})}{N-1}\right).
    \label{eq:density}
\end{align}
As discussed above, the parameter $q_0$ approaches $0$ as $\Np \to \infty$. This suggests that in Model I in which $\kappa$ is constant, the network density converges to $\kappa/{4}$ as $\Np$ (and $N$) grows.

We compare the theoretical and the empirical network density in Fig.~\ref{fig:scaling_fit} (right panels) for CollegeMsg and LondonBike (see Fig.~S6 for the other data sets). For the data sets for which Model I is selected (Interbank, Enron, CollegeMsg and RealityMining), the density monotonically decreases as $N$ increases, approaching the asymptotic value $\widehat{\kappa}/4$ (dashed line). 

For the other data sets (LondonBike and Highschool), on the other hand, the relationship is non-monotonic; density increases with $N$ when $N$ is sufficiently large.
In Model II where $\Np$ is constant, the density can be regarded as a function of $\kappa$, and a shift in $\kappa$ has two effects on the density. First, an increase in $\kappa$ leads the network to be denser because it has a positive impact on the probability of two nodes being connected. Second, an increase in $\kappa$ would cause the number of active nodes $N$ to rise, which has a negative impact on the density. Since there is a finite fraction of inactive nodes when the network is not large enough (i.e., $q_0>0$), the number of active nodes can increase in accordance with a rise in $\kappa$. This increases the denominator of the density by definition, which would lead to a reduction in the theoretical density. Indeed, we find that there exists a threshold of $N$ above which the former effect dominates the latter (Fig.~\ref{fig:scaling_fit}b, \emph{right}).

\subsection{A more general activity distribution}\label{sec:betadist}

The estimation methods we proposed above assume that activity parameters $\{a_i\}$ are distributed uniformly, because in many real-world systems we have no prior knowledge about the activity level of (unobservable) resting nodes. Nevertheless, if we could have further information about the system (in addition to $N$ and $M$), we could also obtain an estimate of the empirical activity distribution that covers the entire set of nodes. 

In this section, we propose a method to estimate activity distribution when the total number of potentially active nodes in the system (i.e., $\Np$) is known. We focus on the systems in which $\Np$ is considered to be constant (i.e., systems for which Model II is selected), namely LondonBike and Highschool, and assume that the true $\Np$ is given by the total number of active nodes of a day.
The implicit assumption here is that nodes that are ready to be active would have at least one temporal edge during a day.
We choose a beta distribution, $\rho(a) = f(a;\alpha,\beta)$ $\equiv\frac{a^{\alpha-1}(1-a)^{\beta-1}}{B(\alpha,\beta)}$ for $a\in [0,1]$, as a general form for the activity distribution. Parameters $\alpha$ and $\beta$ are estimated such that the estimated $\Nph$ matches the empirical counterpart.

A generalized version of the nonlinear regression equation (Eq.~\ref{eq:model2}) is given by (see, Eq.~\eqref{eq:q_0} in Appendix~\ref{sec:analytical})
\begin{align}
    N &= F(M;\Np,\alpha,\beta) \notag \\
      & \equiv \Np \left[ 1- \int da f(a,\alpha,\beta) \right. \notag \\
      & \;\;\;\; \times \left.
      \left(1-\left(\frac{\alpha+\beta}{\alpha}\right)\frac{2Ma}{\Np(\Np-1)}\right)^{\Np-1} \right].\label{eq:model2_gen}
\end{align}
Note that endogenous variable $\kappa$ is now expressed as a function of $M$, taking parameters $\Np$, $\alpha$ and $\beta$ as given: $\kappa(M;\Np,\alpha,\beta)\equiv \left(\frac{\alpha+\beta}{\alpha}\right)^2 \frac{2M}{\Np(\Np-1)}$.

The estimation procedure under a generalized activity distribution is then given by the following four steps:
\begin{enumerate}
    \item For a given combination of $(\alpha,\beta)$, obtain the estimate of $\Np$, denoted by $\Nph(\alpha,\beta)$, by implementing the non-linear least squares on Eq.~\eqref{eq:model2_gen}.
    \item Repeat step 1 for various combinations of $(\alpha,\beta)$.
    \item Find a combination of $(\alpha^*,\beta^*)$ such that  $(\alpha^*,\beta^*) = \argmin_{\alpha,\beta} |\Nph(\alpha,\beta)- N_{\rm p}^{\rm max}|$, where $N_{\rm p}^{\rm max}$ denotes the empirical counterpart of the total number of nodes in the system including temporally resting nodes. 
    \item The estimator of $\Np$ is given by $\Nph(\alpha^*,\beta^*)$.
\end{enumerate}

\begin{figure*}[thb]
    \centering
    \includegraphics[width=11cm]{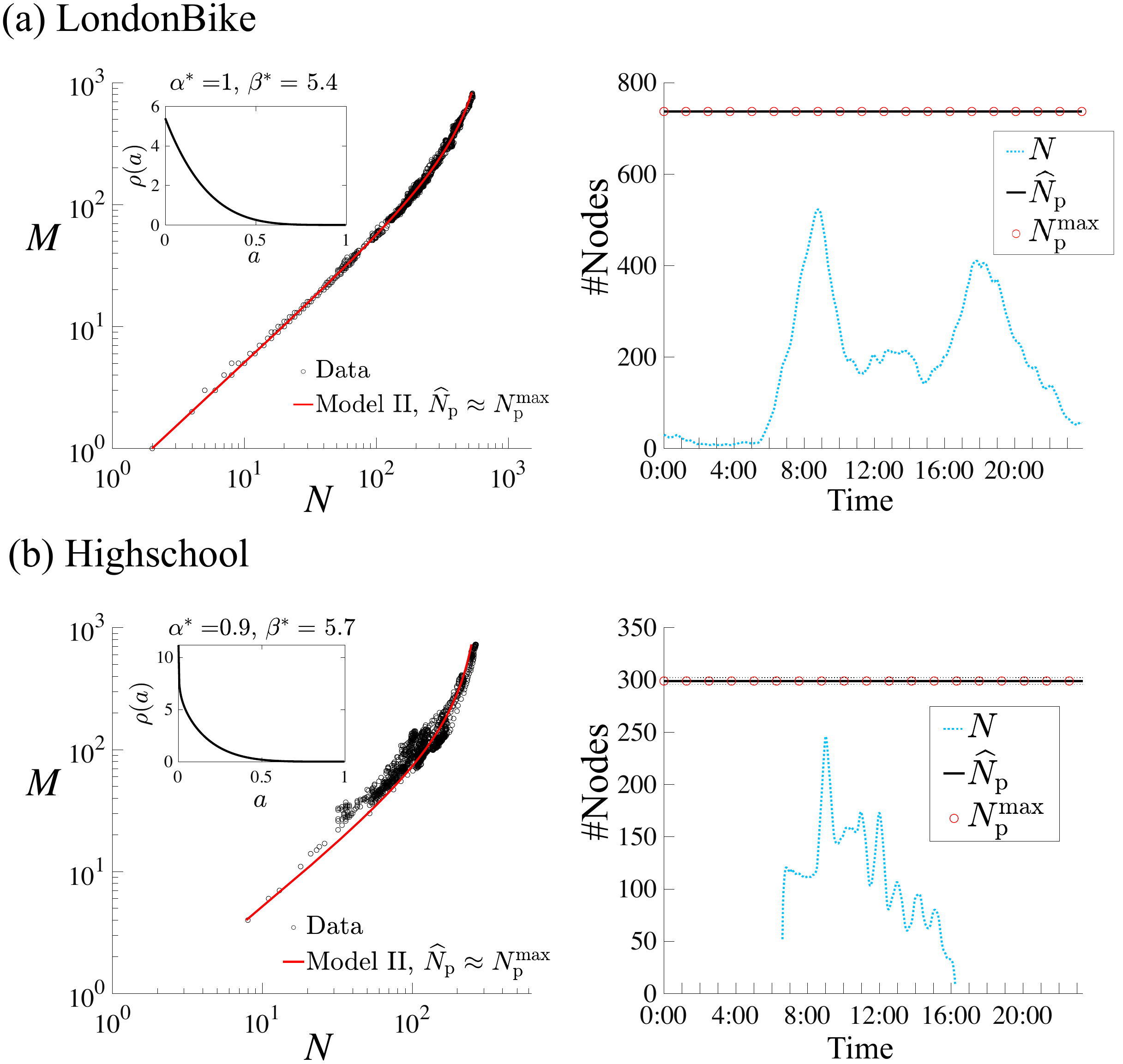}
    \caption{Estimates based on generalized regression equations. In left panels, red line shows the theoretical equation (Eq.~\ref{eq:model2_gen}) with optimised parameters $(\alpha^*,\beta^*)$. \textit{Inset}: Estimated activity distribution $\rho(a)$. In right panels, the number of active nodes $N$ (dotted blue), estimated $\Nph$ (black solid) and the empirical counterpart of $\Np$, denoted by $\Np^{\rm max}$ (red circle), are shown. 5\% confidence interval for $\Nph$ is depicted by black dotted lines.}
    \label{fig:opt_alpha_beta}
\end{figure*}

 The estimation results suggest that the activity distribution is skewed to the left in both data sets (Fig.~\ref{fig:opt_alpha_beta}, \emph{insets}), and the generalized regression equation still well fits the empirical $N$--$M$ curve (Fig.~\ref{fig:opt_alpha_beta}).
We note that while the goodness of fit generally improves due to the introduction of additional parameters (i.e., $\alpha$ and $\beta$), the fitted curve is little affected by the specification of activity distribution (see Figs.~\ref{fig:scaling_fit}b and S2). This suggests that the dynamic hidden variable model well explains the macroscopic fluctuations of empirical networks for alternative specifications of activity distribution.

\section{Discussion}

We proposed a method to identify the source of scaling in temporal networks, namely the dynamical relationship between the numbers of active nodes and edges. Building on a model including both population and activity dynamics, we showed that these two mechanisms are sufficient to explain the two types of scaling observed in real-world systems. The estimating method we developed enables us to compute the parameters for the activity rhythm $\kappa$ and the population size $\Np$ (and thereby the number of resting nodes $\Np-N$). While an observation of $(N,M)$ in a particular snapshot is not sufficient to identify the source of dynamics, a sequence of $N$ and $M$ allows for such an estimation. We apply the method to six empirical data sets, and identify for each the main driving factor responsible for the emergence of scaling.

\add{It should be noted that our proposed framework does not depend on whether the network under study is growing or shrinking. As we already pointed out, the only information needed for the method is a time-variation of $N$ and $M$. Indeed, in many real networks such as the six networks we examined, the size of networks does not exhibit a monotonic behavior but rather non-monotonic shifts, depending on external factors that affect the activity rhythm and/or the population. Thus, the method can reveal the key factor that may lead a network to grow or shrink.}

\add{While our framework is useful for understanding the evolution of temporal networks in any contexts, there remain some issues that need to be addressed in future research. First, our method assumes that there are two types of systems, which are described as Model I (i.e., activity rhythm $\kappa$ is constant and population size $\Np$ is time-varying) and Model II (i.e., population size $\Np$ is constant and activity rhythm $\kappa$ is time-varying). In real-world systems, there may exist an intermediate state in which both the activity rhythm and the population size are evolving with similar time scales. To study those systems, one would need to include additional information other than $N$ and $M$ to inform the model, in order to be able to separate the effects of both mechanisms. 

Second, one key parameter of the model is the distribution of node fitnesses. Currently, we specified this distribution to be either a uniform or a beta distribution, which gives satisfactory estimates of the dynamical parameters. The method would of course yield more accurate estimates if we could incorporate an empirical distribution of fitnesses. However, measuring those is a complicated task: to do so, one needs to observe the activity levels of totally inactive nodes (i.e., nodes without edges), which is paradoxical. The fitness of a node in the model is indeed a rather abstract property, which integrates many realistic characteristics that depend on the context. Such characteristics can also be time dependent.
Third, in the hidden variable model, the structure of the generated network is basically the same as that in a configuration model. Therefore, the model is not sufficient to replicate the empirical structural properties while the aggregate properties are well explained by the model. Explaining the structural and local properties, however, is beyond the scope our paper and should be left for future research. 
}

\appendix

\setcounter{equation}{0}

\renewcommand{\theequation}{A\arabic{equation}}

\section{Data sets}\label{sec:method_data}

The Interbank data set is constructed from bilateral transactions in the online interbank market in Italy between September 4, 2000 and December 31, 2015 (i.e., 3,922 business days). The data is commercially available from e-MID SIM S.p.A. based in Milan, Italy (\verb+http://www.e-mid.it+). From the data we build a temporal network where nodes are banks, with one snapshot per day. For each day, two banks are connected by an edge if a loan is made from a bank to another between 11:00 and 12:00.

The Enron data set is an email-based communication network from the Enron Corporation ~\cite{Klimt2004enron,Leskovec2009enron} collected from May 11, 1999 to June 21, 2002. From the data we build a temporal network where nodes are employees, with one snapshot per day. For each day, two employees are connected by an edge if at least one e-mail has been sent from one employee to the other between 14:00 and 16:00. The data is taken from \cite{network_repository}.

The CollegeMsg data set is an online social network at the University of California, Irvine collected from Mar 23, 2004 to October 26, 2004 ~\cite{Opsahl2008PhysRevLett,Panzarasa09JASIST}. From the data we build a temporal network where nodes are users, with one snapshot per day. For each day, two users are connected if one has sent a private message to the other between 14:00 and 16:00. The data is taken from \cite{SNAP}.

The RealityMining data set is built from the call data from the Reality Commons project ~\cite{Eagle2006PersUbiquitComput} collected from September 24, 2004 to January 7, 2005. From the data we build a temporal network where nodes are individuals, with one snapshot per day. For each day, two individuals are connected if there has been a phone call between them or a voicemail has been left, during the 8:00--12:00 time window. The data is taken from \cite{network_repository}. 

The LondonBike data set describes the trips taken by customers of London Bicycle Sharing Scheme~\cite{LondonBike} collected on January 12, 2016. From the data we build a temporal network where nodes are bike sharing stations, with snapshots every 20 seconds aggregating the data from a 10 minutes sliding time window. For each 10-minutes time interval, two stations are connected if there has been at least one trip between them.

The Highschool data set is a face-to-face contacts network recorded in a high school in France on December 6, 2013, using wearable sensors by the SocioPatterns collaboration ~\cite{Mastrandrea2015PLOS,SocioPatterns}. As in LondonBike, from the data we build a temporal network where nodes are individuals, with snapshots constructed every 20 seconds with a 10 minutes sliding time window. For each 10-minutes time interval, two individuals are connected if they have been at least once in contact.

\section{Full derivation of Eqs.~\eqref{eq:N} and \eqref{eq:M}}
\label{sec:analytical}
In this appendix we show a full derivation of Eqs.~\eqref{eq:N} and \eqref{eq:M}.
The numbers of active nodes $N$ and edges $M$ are expressed as functions of parameters $\kappa$ and $\Np$:
\begin{align}
\begin{cases}
    N &= (1- q_0(\kappa,\Np)) \Np,\\
    M &= \frac{\overline{k}(\kappa, \Np) \Np}{2},
\end{cases}
\label{eq:NM_SI}
\end{align}
where $q_0(\kappa,\Np)$ is the probability of a randomly chosen node being isolated and $\overline{k}(\kappa,\Np)$ denotes the average degree over all the existing nodes including isolated ones.

Given the vector of each node's activity $\vec{a} = (a_1, a_2, \ldots, a_{\Np})$, the probability that node $i$ has degree $k_i$ is written as:
\begin{align}
    g(k_i | \vec{a}) &= \sum_{\vec{c}_i} \left[ \prod_{j \neq i} u(a_i, a_j)^{c_{ij}} (1- u(a_i, a_j))^{1-c_{ij}} \right] \notag \\
    &\times \delta\left(\sum_{j \neq i} c_{ij}, k_i \right),
\label{eq:g_ki}
\end{align}
where $c_{ij} \in \left\{ 0, 1\right\}$ is the $(i,j)$-element of the $\Np\times\Np$ adjacency matrix, whose $i$th column is given by $\vec{c}_i = (c_{1i}, c_{2i}, \ldots, c_{\Np i})^\top$, and function $\delta(x, y)$ denotes the Kronecker delta.

Let us redefine a product term in the square bracket of \eqref{eq:g_ki} as:
\begin{align}
    f_j(c_{ij}; a_i, a_j) &\equiv u(a_i, a_j)^{c_{ij}} (1- u(a_i, a_j))^{1-c_{ij}}.
\label{eq:def_f}
\end{align}
Since $g(k_i | \vec{a})$ is the convolution of $\left\{ f_j(c_{ij}; a_i, a_j) \right\}_j$, its generating function:
\begin{align}
    \hat{g}_i(z | \vec{a}) \equiv \sum_{k_i} z^{k_i} g(k_i | \vec{a})
\end{align}
is decomposed as:
\begin{align}
    \hat{g}_i(z | \vec{a}) = \prod_{j \neq i} \hat{f}_j(z; a_i, a_j),
\end{align}
where $\hat{f}_j$ is the generating function of $f_j(c_{ij}; a_i, a_j)$, given by:
\begin{align}
    \hat{f}_j(z; a_i, a_j) \equiv \sum_{a_{ij}} z^{a_{ij}} f_j(a_{ij}; a_i, a_j).
    \label{eq:def_fhat}
\end{align}

Degree distribution $p(k_i; \kappa,\Np)$ is defined by the probability that node~$i$ has degree $k_i$ and is related to $g(k_i | \vec{a})$ so that:
\begin{align}
    p(k_i ; \kappa,\Np) = \int g(k_i | \vec{a}) \rho(\vec{a}) d \vec{a},
    \label{eq:p_ki}
\end{align}
where we define $\rho(\vec{a}) \equiv \prod_i \rho(a_i)$ and $d \vec{a} \equiv \prod_i da_i$. Therefore, differentiation of $\hat{g}_i(z | \vec{a})$ with respect to $z$ gives the average degree $\overline{k}(\kappa,\Np)$:
\begin{align}
    \overline{k}(\kappa,\Np) &= \sum_{k_i} k_i p(k_i; \Np) \nonumber\\
    &= \sum_{k_i} k_i \int g(k_i | \vec{a})\rho(\vec{a}) d\vec{a} \nonumber\\
    &= \frac{d}{dz} \int \hat{g}_i(z | \vec{a}) \rho(\vec{a}) d \vec{a} \Bigr|_{z=1} \nonumber\\
    &= \frac{d}{dz} \int \rho(a_i) da_i \prod_{j \neq i} \int \hat{f}_j(z; a_i, a_j) \rho(a_j) da_j \Bigr|_{z=1} \nonumber\\
    &= \int \rho(a_i) da_i \frac{d}{dz} \left[ \int \hat{f}(z; a_i, h) \rho(h) dh \right]^{\Np-1} \Bigr|_{z=1} \nonumber\\
    &= (\Np-1)  \int \rho(a_i) d a_i \left[ \int d a \rho(a) \hat{f}(z; a_i, a) \right]^{\Np-2} \notag \\
    &\times \int d a \rho(a) \frac{d}{dz} \hat{f}(z; a_i, a)\Bigr|_{z=1}.
    \label{eq:k1}
\end{align}

From Eqs.~(\ref{eq:def_f}) and (\ref{eq:def_fhat}), we have $\hat{f}(z; a_i, a) = \sum_{c_{ij}} z^{c_{ij}} f(c_{ij}; a_i, a) = (z-1)u(a_i, a) + 1$. It follows that:
\begin{align}
    &\int da \rho(a) \hat{f}(z; a_i, a) = (z-1) \int da \rho(a) u(a_i, a) + 1,\\
    &\int da \rho(a) \frac{d}{dz} \hat{f}(z; a_i, a) = \int da \rho(a) u(a_i, a).
\end{align}


\ From \eqref{eq:p_ki}, the probability of a node being isolated, $q_0(\kappa,\Np) \equiv p(k_i = 0; \kappa,\Np)$, is given by:
\begin{align}
    q_0(\kappa,\Np) 
    &= \int g(k_i =0 | \vec{a})\rho(\vec{a}) d \vec{a} \nonumber\\
    &= \int d a_i \rho(a_i) \left[ 1 - \int u(a_i, a) \rho(a) d a \right]^{\Np-1}.
    \label{eq:q_0}
\end{align}
Then, substituting $\rho(a) = 1$ (i.e., uniform distribution on $[0,1]$) and $u(a, a^\prime) = \kappa a a^\prime$ into Eq.~(\ref{eq:k_avg}) gives:
\begin{align}
    \overline{k}(\kappa,\Np) 
    = \frac{\kappa }{4} (\Np-1).
\end{align}
Similarly, substituting the same conditions into Eq.~(\ref{eq:q_0}) gives:
\begin{align}
    q_0(\kappa,\Np) &= \int_0^1   \left( 1 - \frac{\kappa a_i}{2}  \right)^{\Np-1}d a_i.
\end{align}

By defining a variable as $x \equiv 1 - \frac{\kappa a_i}{2}$, we have:
\begin{align}
    q_0(\kappa,\Np) &=  \frac{2}{\kappa}\int_{1-\frac{\kappa}{2}}^1  x^{\Np-1} dx  \notag \\
    &= \frac{2}{\kappa\Np}\left[1-\left( 1-\frac{\kappa}{2}\right)^{\Np}\right].
\end{align}
Note that $q_0(\kappa,1)=1$ and $\lim_{\Np\to \infty}q_0(\kappa,\Np)=0$.
Combining these results with Eq.~(\ref{eq:NM_SI}), we have:
\begin{align}
\begin{cases}
    N &= \Np \left[ 1-  \frac{2}{\kappa\Np}\left(1-\left( 1-\frac{\kappa}{2}\right)^{\Np}\right) \right],\\
    M &= \frac{1}{8} \kappa\Np(\Np-1).
\end{cases} \label{eq:SI_NM}
\end{align}



\section*{Acknowledgements}
T.K. acknowledges financial support from JSPS KAKENHI Grant nos.~15H05729 and 19H01506.

\section*{Author contributions}
T.K. conceived  and  directed  the  study. T.K. and M.G. defined the model. T.K. performed the analytical calculations and the data analyses. T.K. and M.G. drafted the final manuscript.

\section*{Additional information}

\textbf{Competing financial interests:} The authors declare no competing financial interests.

\clearpage

\setcounter{section}{0}
\setcounter{table}{0}
\setcounter{equation}{0}
\setcounter{figure}{0}
\setcounter{page}{1}
     
\renewcommand{\thetable}{S\arabic{table}}
\renewcommand{\thefigure}{S\arabic{figure}}
\renewcommand{\thesection}{S\arabic{section}}
\renewcommand{\theequation}{S\arabic{equation}}

\begin{widetext}
\fontsize{16pt}{16pt}\selectfont
 Supporting Information:  \\

``Two types of densification scaling in the evolution of temporal networks" \\

\vspace{1cm}
\Large{Teruyoshi Kobayashi and Mathieu G\'enois}
\\
\vspace{1cm}

\begin{figure*}[thb]
    \centering
    \includegraphics[width=16cm]{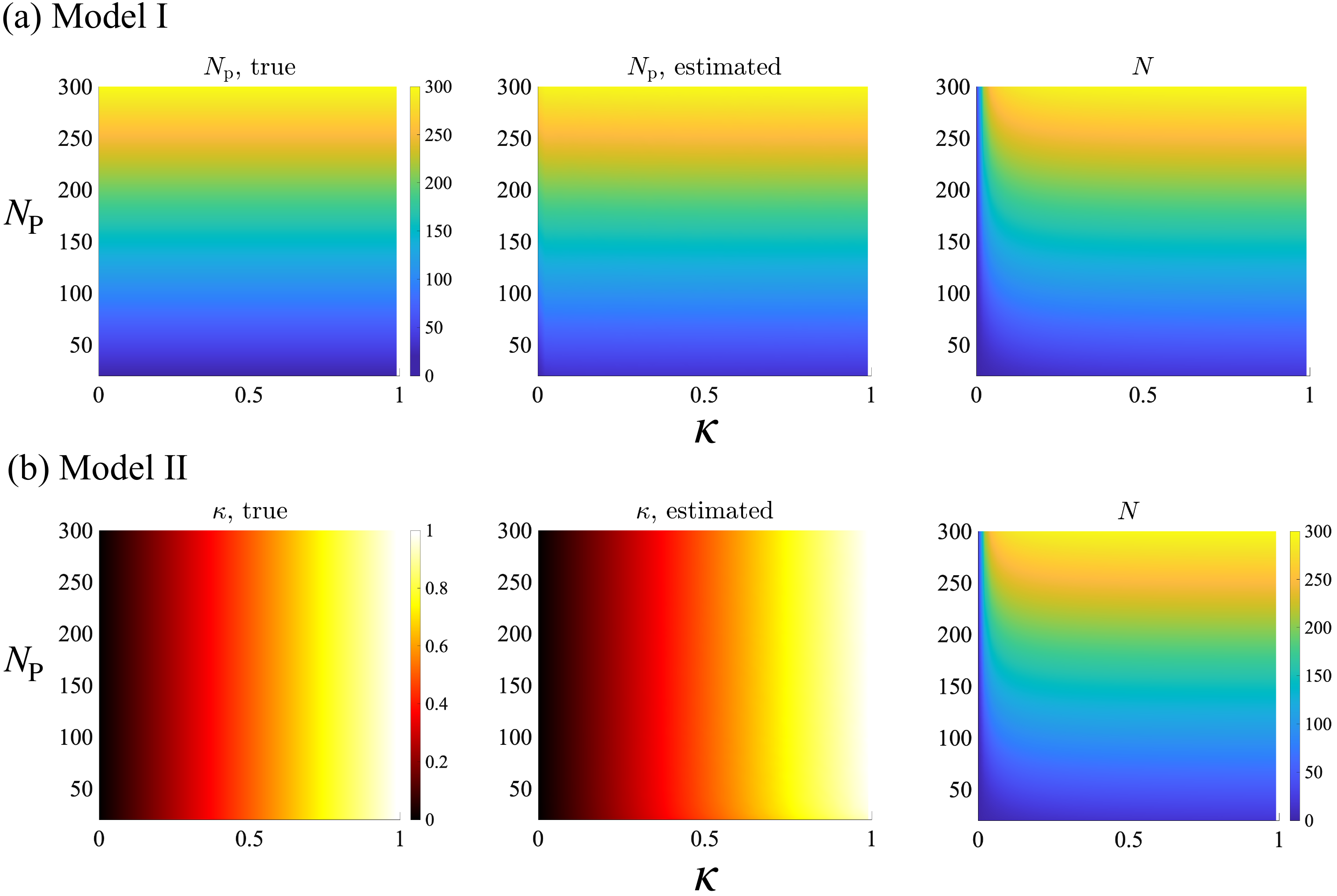}
     \caption{Validation of the proposed estimation method based on synthetic networks. Color denotes the average value over 1,000 runs (color bar in left panel).}
    \label{fig:synthetic_SI}
\end{figure*}

\clearpage

\begin{figure*}[thb]
    \centering
    \includegraphics[width=14cm]{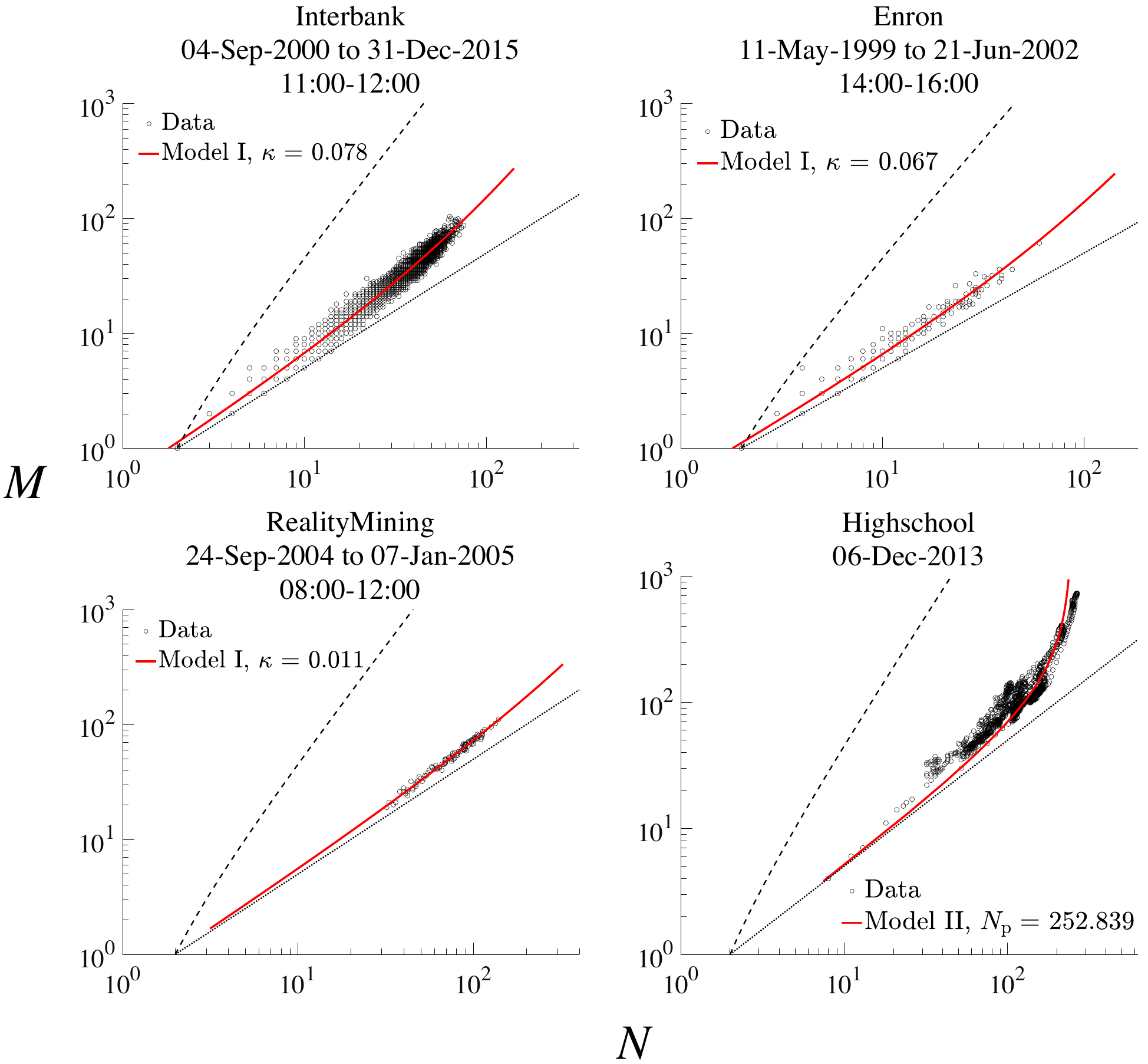}
     \caption{Fitted $M$-$N$ curve. Activity is uniformly distributed: $a\in [0,1]$.}
    \label{fig:scaling_fit_SI}
\end{figure*}

\clearpage

\begin{figure*}[thb]
    \centering
    \includegraphics[width=14cm]{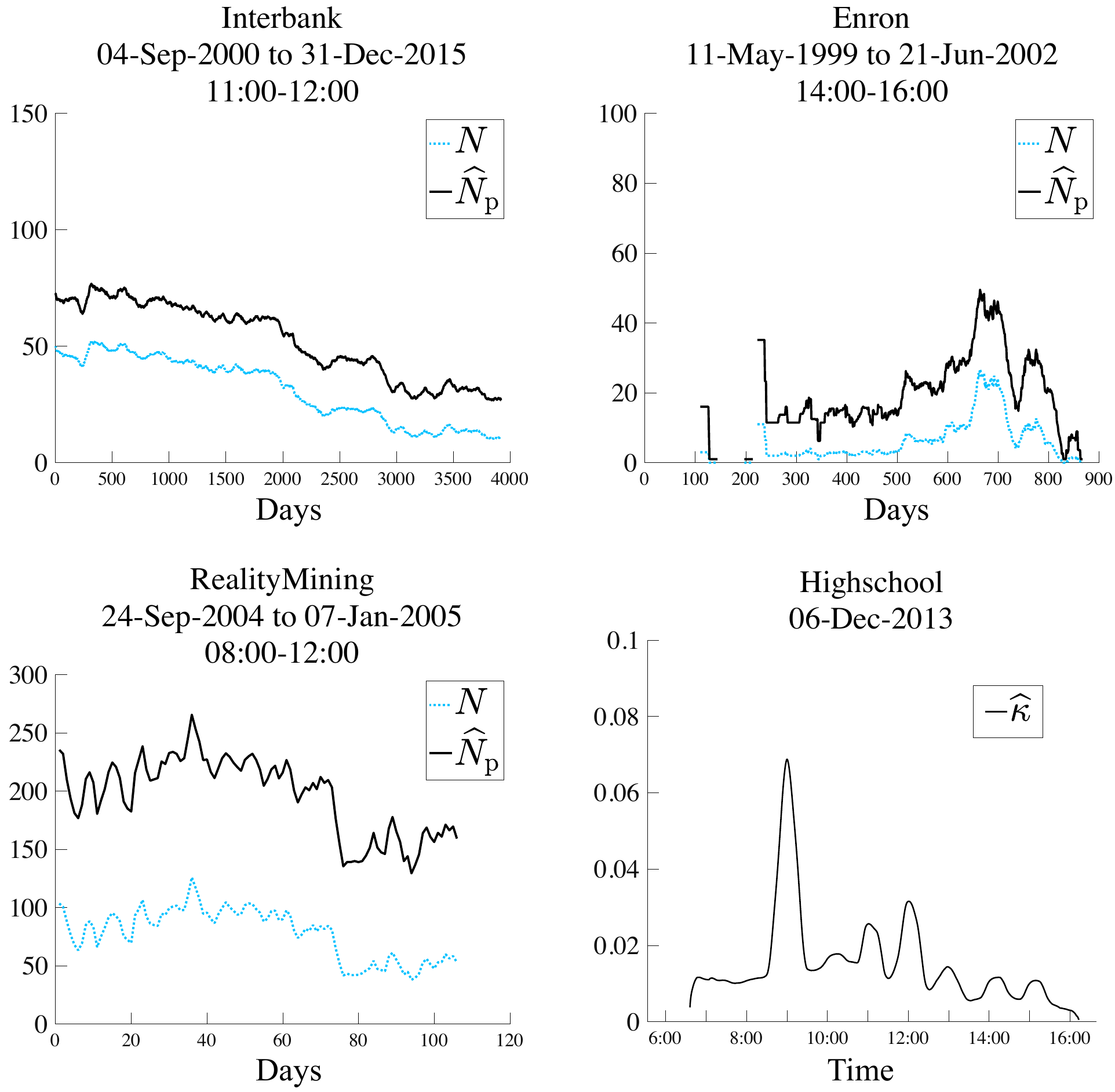}
     \caption{Estimated time-varying parameters. Activity is uniformly distributed: $a\in [0,1]$. Model I: Interbank, Enron and RealityMining. Model II: Highschool.}
    \label{fig:estimate_param_SI}
\end{figure*}

\clearpage
\begin{figure*}[thb]
    \centering
    \includegraphics[width=16cm]{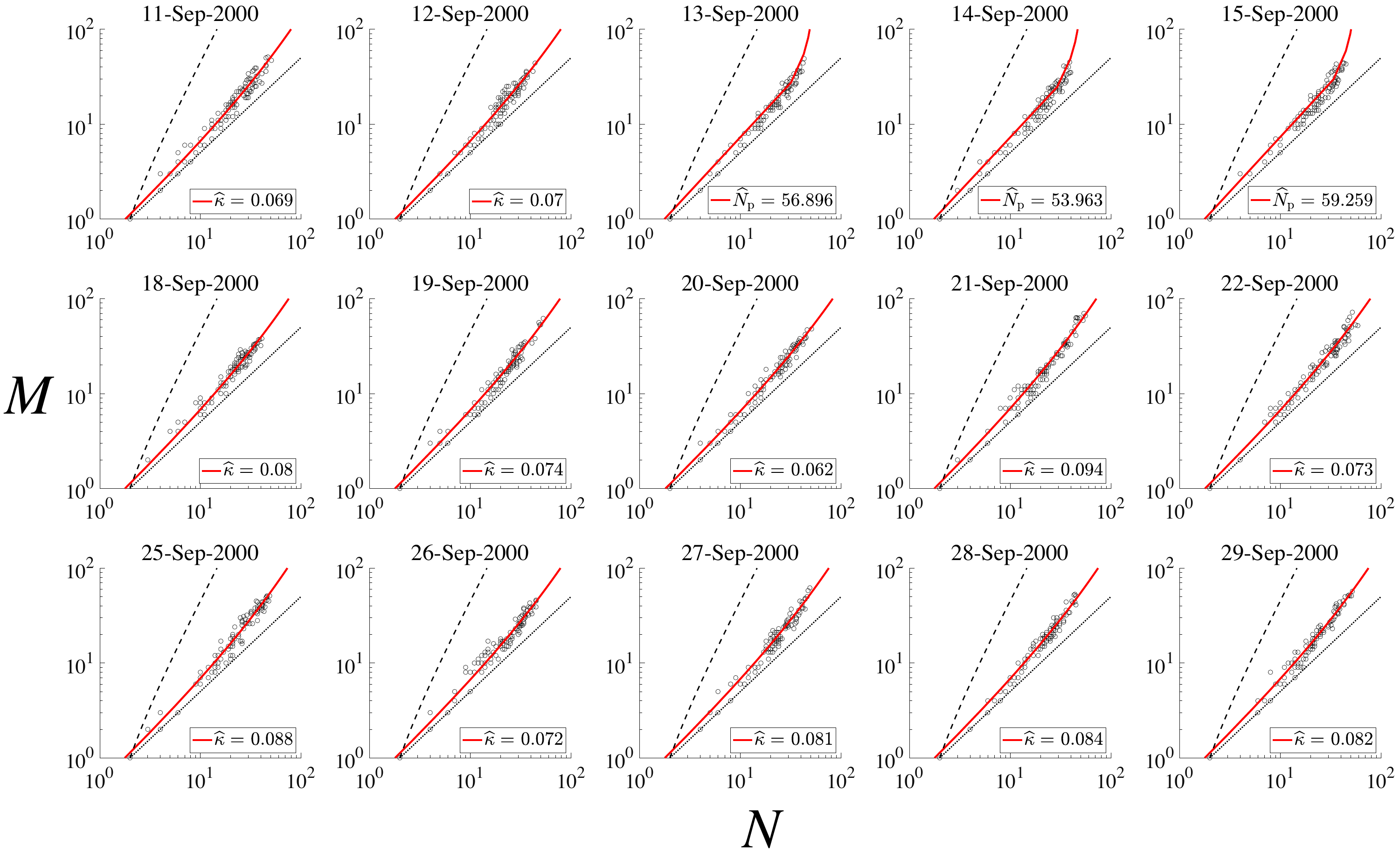}
     \caption{Model fit to the intra-day networks in the Interbank data set. Red line denotes theoretical values indicated by the selected model. For each day, each dot represents a snapshot network of a 20-minute time window. Snapshots are created every 5 minutes. Model I (Model II) is selected for panels showing estimated ${\kappa}$ ($\Np$). Model II is selected for September 13, 14 and 15 while Model I is selected for the other 12 days.}
    \label{fig:emid_eachday_SI}
\end{figure*}

\clearpage
\begin{figure*}[thb]
    \centering
    \includegraphics[width=16cm]{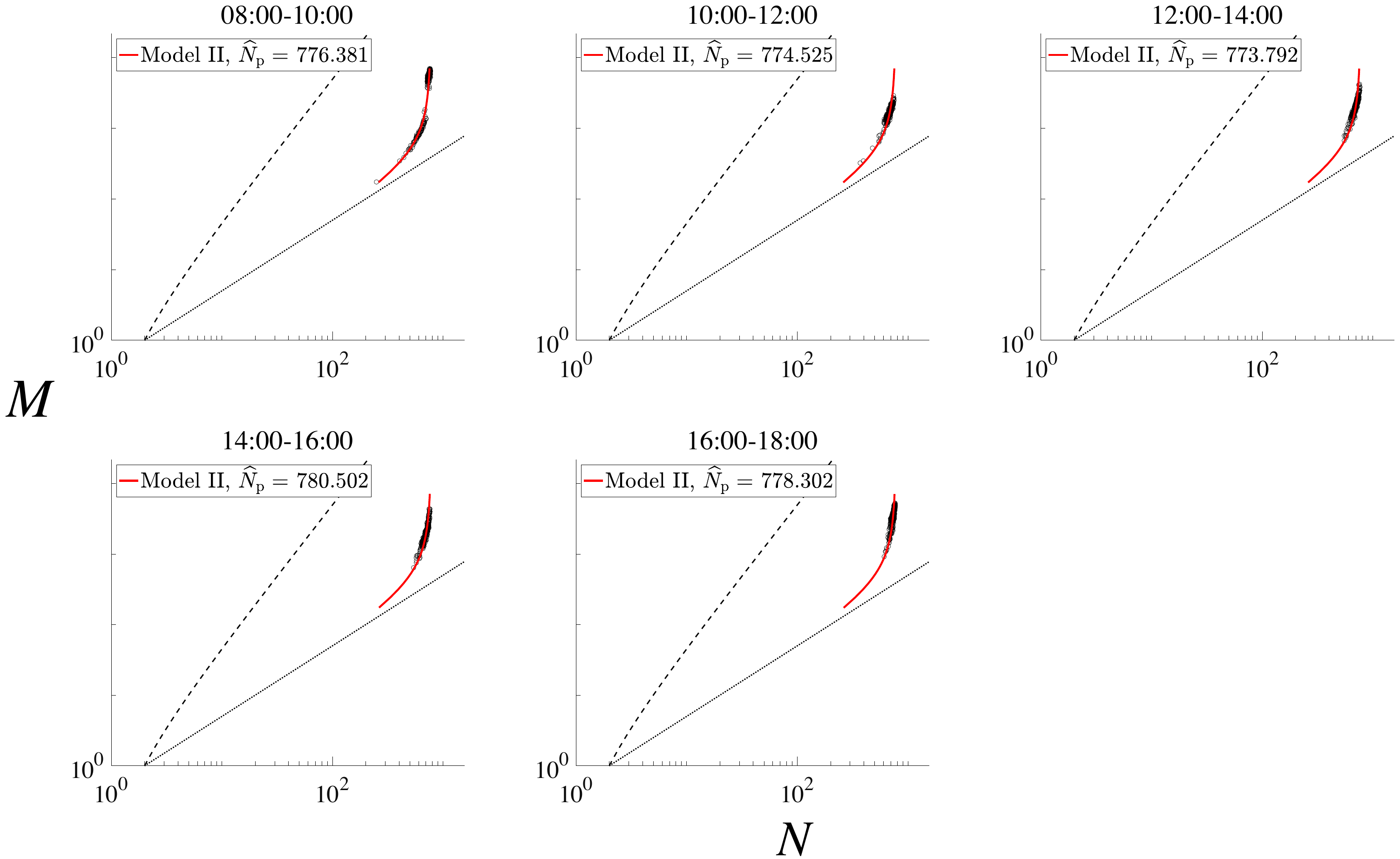}
     \caption{Model fit for the LondonBike dataset in each time bin. The data period ranges from January 10, 2016 to August 31, 2016 (235 days). Red line denotes theoretical values indicated by the selected model. Model II is selected for all time bins, indicating that the number of existing nodes is constant.}
    \label{fig:LondonBike_eachtimebin_SI}
\end{figure*}

\clearpage
\begin{figure*}[thb]
    \centering
    \includegraphics[width=14cm]{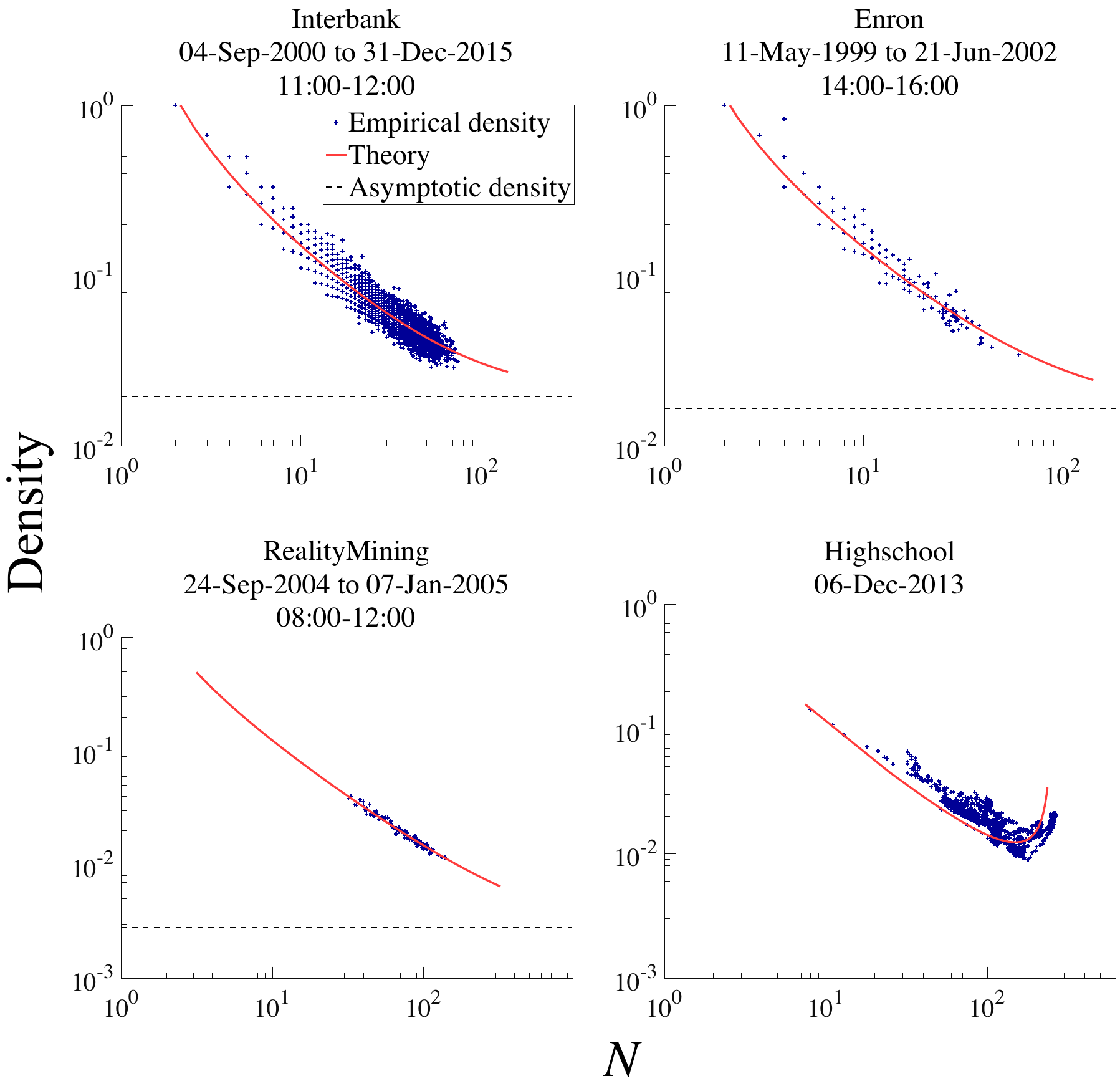}
     \caption{Empirical and theoretical density. Activity is uniformly distributed: $a\in [0,1]$. Model I: Interbank, Enron and RealityMining. Model II: Highschool.}
    \label{fig:density_SI}
\end{figure*}

\clearpage

\end{widetext}

\end{document}